\documentclass[11pt,letter,defpaper]{article}
\usepackage{psfig}
\usepackage{defpaper}
\usepackage{latexsym}
\include{psfig}

\newcommand{\usepdf}[1]{}
\newcommand{\ifpdf}[1]{}

\newcommand{\ifps}[1]{#1}

\eatreminders
 
\newcommand{\ccm}[1]{client cache middle-ware}

\newcommand{\incremental}{{\sf Incremental}}

\newcommand{\nocache}{{\sf NoCache}}
\newcommand{\infcache}{{\sf InfCache}}
\newcommand{\adms}{{\sf LCS/LRU}}

\title{
Don't Trash your Intermediate Results, Cache 'em\thanks{
Work partly funded by a Govt. of India DST grant and a UPP grant from IBM.
}}
\author{
{\bf Prasan Roy}
\thanks{Work partly funded by an IBM PhD fellowship.} $^1$
\hspace*{1cm}
{\bf Krithi Ramamritham}\thanks{Also affiliated with Univ. of
Mass. Amherst.} $^1$
\and
{\bf S. Seshadri}\thanks{Work primarily done while at IIT Bombay.}  $^2$
\hspace*{1cm}
{\bf Pradeep Shenoy}$^2$
\hspace*{1cm}
{\bf S. Sudarshan}$^1$
\and 
$^1$ IIT Bombay\\
\{prasan,krithi,sudarsha\}@cse.iitb.ernet.in \\
$^2$ Bell Laboratories, Murray Hill \\
\{seshadri,pshenoy\}@research.bell-labs.com
}
\date{}
\begin{document}

\eat{
{\Large
\hspace*{\fill} Paper Number: {\bf 193} \\
\hspace*{\fill} Category: {\bf Research} \\
}
\\
{\large
\begin{center}
{\Large\bf Don't Trash your Intermediate Results, Cache 'em } \\
~\\
{\bf Prasan Roy},  {\bf Krithi Ramamritham}, {\bf S. Seshadri} \\
{\bf Pradeep Shenoy} and {\bf S. Sudarshan}\\
\end{center}
~\\
\begin{tabular}{ll}
Contact author &  Krithi Ramamritham \\
& krithi@cs.umass.edu\\
& Computer Science Department \\
& University of Massachusetts \\
& Amherst, Mass. 01003-4610 \\
& (413) 545-0196 (office) \\
& (413) 545-1249 (fax) \\
\end{tabular}
}
}

\thispagestyle{empty}
\pagebreak
\setcounter{page}{1}

\maketitle

\eat{
\begin{abstract}
In data warehouse and data mart systems, queries often take a long
time to execute due to their complex nature.  Query response times can
be greatly improved by caching final/intermediate results of previous
queries, and using them to answer later queries. In this paper we
describe a caching system called Exchequer which incorporates several
novel features including optimization aware cache maintenance and the use
of a cache aware optimizer. In contrast, in existing work, the module
that makes cost-benefit decisions is part of the cache manager and
works independent of the optimizer which essentially reconsiders these
decisions while finding the best plan for a query. In our work, the
optimizer takes the decisions for the cache manager.  Furthermore,
existing approaches are either restricted to cube (slice/point)
queries, or  cache just the query results. On the other hand, our work is extensible
and in fact presents a data-model independent framework and algorithm.
Our experimental results attest to the efficacy of our cache
management techniques and show that over a wide range of parameters
(a) Exchequer's query response times are lower by more than 30\%
compared to the best performing competitor, and (b) Exchequer can
deliver the same response time as its competitor with just one tenth
of the cache size.
\end{abstract}
}

\begin{abstract}
In data warehouse and data mart systems, queries often take a long
time to execute due to their complex nature.  Query response times can
be greatly improved by caching final/intermediate results of previous
queries, and using them to answer later queries. An automatic
caching system that makes intelligent decisions on what results to cache
would be an important step towards knobs-free operation of a 
database system, essentially a step towards a 
database system that auto-tunes itself. 

In this paper we describe an automatic query caching system called 
{\em Exchequer} which is closely coupled with the optimizer to ensure
that the caching system and the optimizer make mutually consistent decisions. 
In contrast, in existing work, such a close coupling is absent.
We present a cache management/replacement algorithm which we call
\incremental.
Furthermore, existing approaches are either restricted to 
cube (slice/point) queries, or to caching just the final query results. 
On the other hand, our work is extensible and in fact presents a 
data-model independent framework and algorithm.
Our experimental results attest to the efficacy of our cache
management techniques and show that over a wide range of parameters
(a) Exchequer's query response times are lower by more than {\em 30\%}
compared to the best performing competitor, and (b) Exchequer can
deliver the same response time as its competitor with just {\em one
tenth} of the cache size.
\end{abstract}

\sections{Introduction}
\label{sec:intro}

Data warehouses and On-Line Analytical Processing (OLAP) systems are
becoming increasingly important parts of data analysis.  The typical
processing time of decision support and OLAP queries range from minutes
to hours. This is due to the nature of complex queries used for
decision making. Data warehouses act as mediators between data sources
and data consumers.  Various OLAP, data mining, querying and reporting
tools access data at the central warehouse.  The aim of our work is to
improve the query response time by caching the (final as well as
intermediate) results produced during query processing.

In a traditional database engine, every query is processed independently. 
In decision support applications, the query workload, consisting of a 
sequence of queries, often has overlapping expressions between queries.
A natural way to improve performance is to cache and reuse results of
previous query results when evaluating later queries.
The unit of caching need not just be final results of 
queries --- intermediate results of queries may also be cached.

We use the term {\em query caching} in this paper to mean caching 
of final and/or intermediate results of queries.
Query caching differs from standard page caching in three major ways.
\begin{enumerate}

\item \label{point:cost}
In contrast with page replacement schemes, where  recency of
usage is often sufficient,  other factors such as size and cost of 
computation of  results must be considered.

\item \label{point:opt}
The query optimizer must be extended to consider the possibility of 
using cached results of previous queries. 

\item \label{point:benefit}
In traditional page caching, the pages cached are independent
of each other in that the presence (or absence) of a page in the cache
does not affect the benefit of caching another page for
a given workload. In contrast, a given query may be computable from 
any one of several query results (or intermediate query results).
Therefore, the benefit (to the given query) of caching a particular
query result depends on what else is present in the cache. 

\end{enumerate}
%

There has been some work on caching of query results in the recent past.
We discuss related work in more detail in Section~\ref{sec:relwork}, but
outline the main differences of our approach from directly related prior 
work here.  Earlier work on caching has been for specialized applications (e.g.
data cubes \cite{des98:chunks,kot99:dynamat,sch99:dyncache}, 
or just selections \cite{dar96:semcache,basu96:predcache}),
or does not take caching of intermediate results into account
(\cite{sch96:watchman}),
or has relatively simple cache replacement algorithms,
which do not take into account the fact that the benefit of a cached 
result may depend on what else is in the cache
(\cite{rou:edbt94}, which handles only select-project-join
queries).

There has been work on the related area of materialized view/index selection,
which can be viewed as a static version of caching, where the cache
contents do not vary
(e.g., see \cite{rss96:matview,lqa97:phys,gupta97:viewsel} for general views,
\cite{venky:sigmod96,venky:index,wisc:matview} for data cubes,
and \cite{sn97:indexsel} for index selection).
Techniques for materialized view/index selection
use sophisticated ways of deciding what to materialize,
where the computation of the benefit of materializing a view takes into
account what other views are materialized.
However, this body of work does not consider dynamic changes, and ignores
the cost of materializing the selected views.
The major disadvantage of static cache contents is that it cannot cater to 
changing workloads. 
The data access patterns of the queries cannot be expected to be static. 
To answer all types of queries efficiently, we need to dynamically change 
the cache contents.
Another related area is multiquery optimization, where, e.g.,
the work of \cite{rssb00:mqo} takes the cost of materializing the selected
views, but still makes a static decision on what to materialize
based on a fixed workload. 
In this paper, we study how to adapt the sophisticated 
techniques proposed for materialized view selection,
to solve the problem of query caching.
A first cut approach could be to simply run a view selection algorithm 
periodically.
However, doing so naively (a) could increase the cost of materializing the selected views, and 
(b) would not be able to react quickly to changes in the query load.
Hence we propose and evaluate more sophisticated alternatives, 
and show that dealing with the dynamics of query workloads is not only
possible  but also necessary for improved performance.  
The techniques presented in this paper form part of the
Exchequer\footnote{Efficiently eXploiting caCHEd QUEry Results} query 
caching system, which we are currently implementing.
\eat{being built at IIT Bombay.}

The contributions of this paper are as follows:
\begin{itemize}
\item We show how to use materialized view/index selection techniques 
for short to medium term caching.  
To achieve this we have developed efficient techniques for
exploiting sharing opportunities that arise when several related 
queries follow one another, 
and it is useful to cache their results even if they are not part of a 
long-term trend.

\item

\eat{
In particular, we address the 
issue of {\em when} and {\em how} to materialize results selected
for caching.  Wherever possible,  slack time can be used to 
materialize the selected results before any query arrives, but 
}
In general we have to interleave query execution with materializing 
selected results. Doing this efficiently requires a careful analysis 
of the costs and benefits of forcing a query to execute in
such a way as to  materialize a	selected result.

\item

We cache intermediate as well as final results. Intermediate
results, in particular, require sophisticated handling,  since 
caching decisions are typically made based on usage rates;
usage rates of intermediate results are dependent 
on what is in the cache, and techniques based only on usage
rates would be biased against results that happen not to be 
currently in the cache.  

Our caching algorithms exploit sophisticated 
techniques for deciding what to cache, taking into account what other 
results are cached. Specifically, using incremental techniques developed by us in \cite{rssb00:mqo},
	        we are able to efficiently compute benefits of
                final/intermediate results that are candidates for caching.

\item We use an AND-OR DAG representation of queries and cached results,
	as in Volcano \cite{gra:vol}: 
	\begin{itemize}
	\item The representation is extensible to new operations, 
	unlike much of the prior work on caching, and efficiently 
	encodes alternative ways of evaluating queries. In particular,
	therefore, our algorithms can handle any SQL query including nested
	queries.  To the best of our knowledge, no other
	caching technique is capable of handling such a general class 
	of queries. 

	\item The representation allows the optimizer to efficiently
		take into consideration the use of cached results.

	\item Our DAG representation also includes physical properties such
		as sort orders, and the presence of indexes.  Thus we 
		are able to consider caching indices
		constructed on the fly in the same way
                as we consider caching of intermediate query results.
	\end{itemize}



\item We present an implementation-based performance study that clearly 
demonstrates the benefits of our approach over earlier approaches.
Our study shows that
\begin{itemize}
\item
  intelligent caching can be done fast enough to be practical, and leads
  to huge overall savings, and 
\item
our approach performs better than previously proposed algorithms. For
a given cache size, our response times are
at least  30\% lower than the nearest competitor. Furthermore, we can
deliver the same response times as our nearest competitor with just
one tenth of the cache size.
\end{itemize}
Overall, our experimental results indicate that storing intermediate
results is important, and using sophisticated caching techniques
(which invest extra effort in careful transient view selection) is
well worth it.

\end{itemize}

The rest of the paper is organized as follows.  We start with a brief
description of the architecture of the Exchequer query caching system, 
and an overview of our dynamic query caching algorithms in 
Section \ref{sec:exchequer}.
Section~\ref{sec:dagrep} covers background material regarding
the DAG (directed acyclic graph) representations that underlie our algorithm implementations.
Details of the adaptive caching strategies, including when to 
materialize a result chosen for caching are presented in 
Section \ref{sec:algos}.
Results of experimental evaluation of the proposed algorithms are
discussed in Section \ref{sec:perf}. 
Related work is covered in detail in Section~\ref{sec:relwork}, and
conclusions and future work are outlined in Section~ \ref{sec:concl}.


\sections{Overview of Exchequer's Query Caching Architecture and Algorithms}
\label{sec:exchequer}

\begin{figure*}
\centerline{
\ifpdf{
\mbox{\pdfimage width 6.0in {figures/model.pdf} \relax}
}
\ifps{
\psfig{file=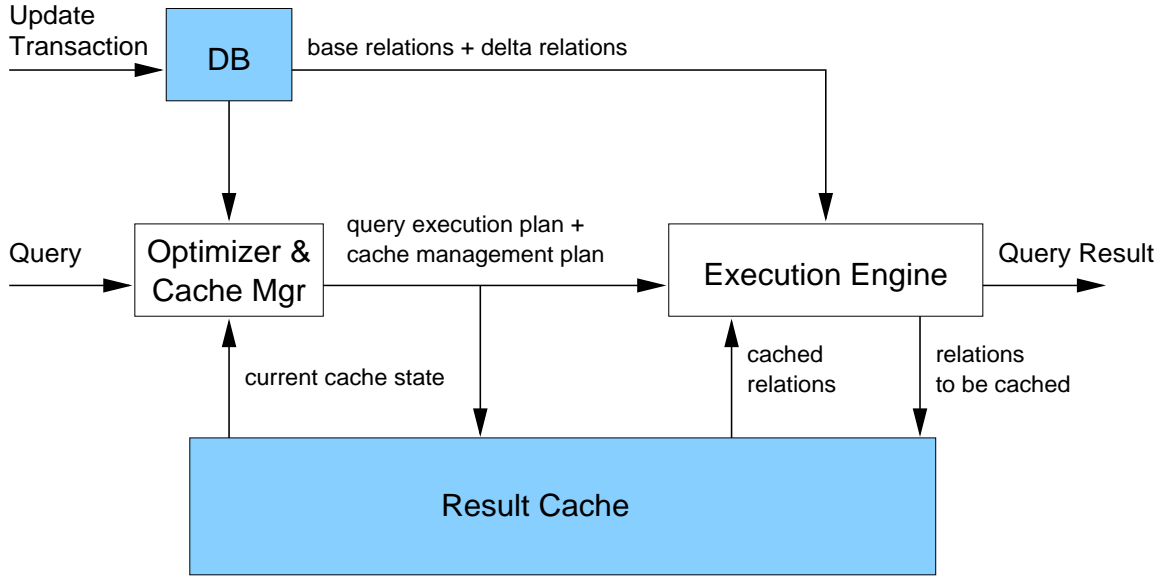,width=6in}
}
}
\caption{Architecture of the Exchequer System}
\label{fig:model}
\end{figure*}

The caching algorithms presented in this paper form part of the 
Exchequer query caching system.
The architecture of the Exchequer system is portrayed in 
Figure~\ref{fig:model}. 


%

As shown, the optimizer optimizes an incoming 
query based on the current cache state. We have integrated the
cache manager and the optimizer and this tight coupling is
demonstrated in Figure~\ref{fig:model}. The cache manager
decides which intermediate results to cache and which cached
results to evict based on the workload (which depends
on the sequence of queries in the past). Conceptually, as the
architecture shows, for each query, the optimizer cum cache manager
produces the query execution plan as well as the cache
management plan (which describes changes to the cache state if any).
A query execution plan may refer to cached relations
(since the optimizer is cache-aware) which are obtained from the query cache 
by the execution engine when required. 
In addition, new intermediate results produced by the query may be cached
on the advise of the cache manager.


In addition to the above functionality, a caching system should also
support invalidation or refresh of cached results in the face of
updates to the underlying database.  Exchequer does support this 
functionality.  In this paper, however, we will confine our attention 
only to the issue of efficient query processing, ignoring updates.  
Data Warehouses are an example of an application where the cache 
replacement algorithm can ignore updates, since updates happen only 
periodically (once a day or even once a week).

The techniques needed  (a) for managing the cache contents intelligently as 
queries arrive and (b) for performing
query optimization exploiting the cache contents, form the crux of 
this paper.  The main question related to
intelligent cache management and cache-aware optimization is:
\begin{enumerate}
\item
How do we dynamically characterize a changing workload  and
accordingly decide the caching 
and query optimization 
strategies so as to minimize the overall response time for all the queries?
\end{enumerate}
Answering this question presents several subproblems:
\begin{itemize}
\item 
Characterizing the dynamically changing 
query workload so as to construct a
{\em representative set of queries}, denoted, henceforth, by {\em repset}, 
to make dynamic caching decisions with.

Our  model for changing workloads is based on a 
representative set of the past N distinct queries
with attached weights which are related to their recency of occurrence.
The cache manager (incrementally) maintains the consolidated DAG
of these queries. We explain this model in more detail
in Section~\ref{sec:algos}.
\item 
Given the dynamic characteristics of the workload, dynamically  adapting the
contents of the cache to suit the needs of the workload.

The decisions concerning what to cache are based on the computation of
the benefits accruing from caching a particular set of materialized
views (specifically, final query results as well as intermediate
results).
The cache manager adapts the cache contents to both long term variations
in the workload and short term variations in the workload. 

One of the goals of using a multi-query optimizer is to reuse a set of
intermediate results (across queries) so as to reduce the overall response
time for a set of queries. If these intermediate results are cached, this
is precisely what is needed for the caching problem also, at a
certain point in time.  Hence we adapt a multi-query 
optimizer developed earlier~\cite{rssb00:mqo} 
to decide what should be cached. 

\item
Optimizing the execution of a set of queries in a
cache-cognizant fashion.

A by-product of the
multi-query optimizer is an optimal plan for a set of queries given a
set of materialized views. This proves useful for cache-cognizant
optimization as well.

\end{itemize}

In the next section we discuss the AND--OR DAG representation of queries and
then show how this representation aids in solving the above subproblems.

\sections{DAG Representation of Queries}
\label{sec:dagrep}

An AND--OR DAG is a directed acyclic graph whose nodes can be divided
into AND-nodes and OR-nodes;  the AND-nodes have only OR-nodes 
as children and OR-nodes have only AND-nodes as children. 

An AND-node in the AND-OR DAG corresponds to an algebraic operation, 
such as the join operation ($\Join$) or a select operation ($\sigma$).
It represents the expression defined by the operation and its inputs. 
Hereafter, we refer to the AND-nodes as {\em operation nodes}.
An OR-node in the AND-OR DAG represents a set of logical 
expressions that generate the same result set; the set of such expressions
is defined by the children AND nodes of the OR node, and their inputs.
We shall refer to the OR-nodes as {\em equivalence nodes} henceforth.

\begin{figure*}
\centerline{
\ifpdf{
\mbox{\pdfimage width 6.0in {figures/dagex.pdf} \relax}
}
\ifps{
\psfig{file=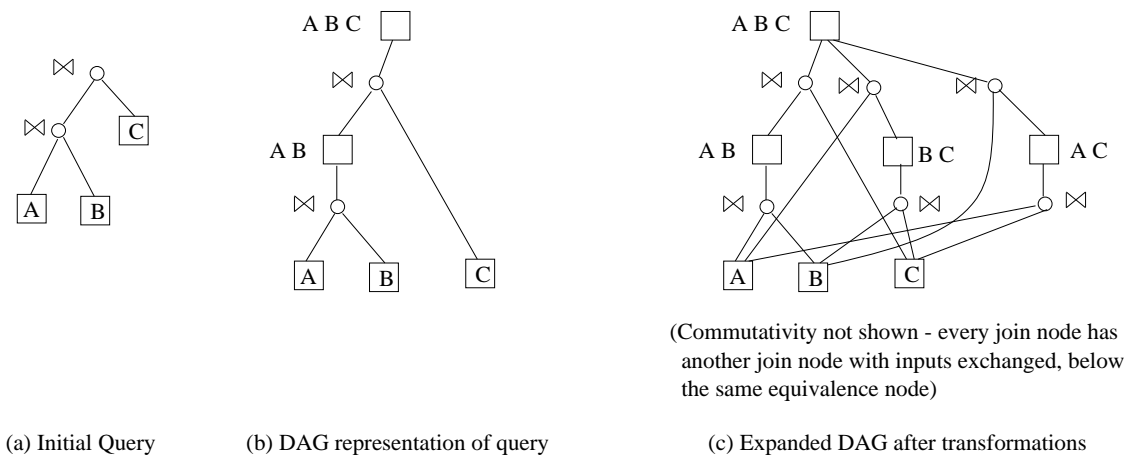,width=6in}
}
}
\caption{Initial Query and DAG Representations}
\label{fig:vol:dag}
\end{figure*}

\subsections{Representing a Single Query}

The given query tree is initially represented directly in the AND-OR DAG 
formulation.
For example, the query tree of Figure~\ref{fig:vol:dag}(a) is initially
represented in the AND-OR DAG formulation, as shown 
in Figure~\ref{fig:vol:dag}(b).
Equivalence nodes (OR-nodes) are shown as boxes, while
operation nodes (AND-nodes) are shown as circles.

The initial AND-OR DAG is then expanded by applying all possible 
transformations on every node of the initial query DAG representing
the given set of queries. 
Suppose the only transformations possible are join associativity
and commutativity.
Then the plans $A \Join (B \Join C)$ and $(A \Join C) \Join B$,
as well as several plans equivalent to these modulo commutativity
can be obtained by transformations on the initial AND-OR-DAG
of Figure~\ref{fig:vol:dag}(b).
These are represented in the DAG shown in Figure~\ref{fig:vol:dag}(c).
We shall refer to the DAG after all transformations have been applied
as the {\em expanded DAG}.
Note that the expanded DAG has exactly one equivalence node for every 
subset of $\{A, B, C\}$; the node represents all ways of computing 
the joins of the relations in that subset.  

\subsections{Representing Sets of Queries in a DAG}
\label{sec:dagrep:queryset}

Given a {\em repset}, the representative set of queries used to make dynamic
caching decisions (how repset is chosen is described later,
in Section \ref{sec:dagrep}), we represent the queries in repset in a 
single {\em consolidated} DAG structure. 
Since repset evolves with time, we need to insert and
delete queries from the DAG.  Since parts of the DAG may be shared by
multiple queries, deletion of intermediate nodes of the DAG is done 
by a reference counting mechanism,

Queries are inserted into the DAG structure one at a time.
When a query is inserted, equivalence nodes and operation nodes are created
for each of the operations in its initial query tree. 
Some of the subexpressions of the initial query tree may be equivalent to
expressions already in the DAG.  Further, subexpressions of a query 
may be equivalent to each other, even if syntactically different.
For example, suppose a query contains a subexpression that is logically
equivalent to, but syntactically different from another subexpression (e.g., $(A \Join B) \Join C$, and $A \Join (B \Join C)$).
Before the second subexpression is expanded, the DAG would contain 
two different equivalence nodes representing the two subexpressions.  
We modify the Volcano DAG generation algorithm so that whenever
it finds nodes to be equivalent (in the above example, after applying 
join associativity) it {\em unifies} the nodes, replacing them by 
a single equivalence node.
The Volcano optimizer \cite{gra:vol} already has a hashing-based scheme to 
efficiently detect repeated expressions, thereby avoiding creation of 
new nodes that are 
equivalent to existing nodes.  Our extension
additionally unifies existing 
equivalent nodes.

Another extension is to detect and handle {\em subsumption} derivations.
For example, suppose two subexpressions $e1$: $\sigma_{A<5}(E)$ and
$e2$: $\sigma_{A<10}(E)$ appear in the query.
The result of $e1$ can be obtained from the result of $e2$ by
an additional selection, i.e.,
$\sigma_{A<5}(E) \equiv \sigma_{A<5}(\sigma_{A<10}(E))$.
To represent this possibility we add an extra operation node 
$\sigma_{A<5}$ in the DAG, between $e1$ and $e2$.
Similarly, given $e3$: $\sigma_{A=5}(E)$ and $e4$: $ \sigma_{A=10}(E)$,
we can introduce a new equivalence node $e5$: $\sigma_{A=5 \vee A=10}(E)$ and 
add new derivations of $e3$ and $e4$ from $e5$.
The new node represents the sharing of accesses between the two selection.
In general, given a number of selections on an expression $E$, we 
create a single new node representing the disjunction of all the
selection conditions.
Similar derivations also help with aggregations.
For example, if we have $e6$: $_{dno} {\cal G}_{sum(Sal)} (E) $ and
$e7$: $_{age} {\cal G}_{sum(Sal)} (E) $, we can introduce a new equivalence
node $e8$: $_{dno,age} {\cal G}_{sum(Sal)} (E) $ and add derivations of
$e6$ and $e7$ from equivalence node $e8$ by further groupbys on $dno$ and 
$age$.

More details of unification and subsumption derivations, which we developed
and used initially for multi-query optimization, can be found 
in \cite{rssb00:mqo}.

\subsections{Physical Properties}
\label{ssec:phys:dag}

It is straightforward to refine the above AND-OR DAG representation to
represent {\em physical properties} \cite{gra:vol}, such as sort order,
that do not form part of the logical data model, and obtain 
a physical AND-OR DAG
\footnote{For example, an equivalence node is refined to multiple
physical equivalence nodes, one per required physical property, in the
physical AND-OR DAG.}.  
We also model the presence of an index on a result as a physical property
of the result.   Physical properties of intermediate results
are important, since e.g., if an intermediate result is sorted on a
join attribute, the join cost can potentially be reduced by using a
merge join.  This also holds true of intermediate results that are
materialized and shared.  Our implementation of the AND-OR DAG model
for caching purposes and  our search algorithms indeed handle 
physical properties.  However, for pedagogical reasons, we do not
explicitly consider physical properties further.


\sections{Details of the Exchequer System}
\label{sec:algos}

We first discuss how optimization of a query is carried out in a cache
cognizant fashion.  To select results for caching, we need to be able 
to predict the future, based on the past.  We therefore describe
a model based on representative sets of queries, which permits us 
to select final and intermediate results.
We then describe a greedy algorithm for selecting results for caching,
and consider optimizations of the algorithm.

\subsections{Cache-Aware Query Optimization}
\label{sec:volc:mat}

We now consider how to perform 
cache-aware optimization. We first describe the basic Volcano 
optimization algorithm, and then
extend it to handle cached results.

The Volcano optimization algorithm finds the best plan for each node 
of the expanded DAG by performing a depth first traversal of the DAG.
Costs are defined for operation and equivalence nodes.
The cost of an operation node is $o$ is defined as follows:\\
\hspace*{1cm}
$cost(o) = $ cost of executing $(o)$ + $\Sigma_{e_i \in children(o)}
cost(e_i)$\\
The children of $o$ (if any) are equivalence nodes.
The cost of an equivalence node $e$ is given as \\
\hspace*{1cm}
$cost(e) = min \{ cost(o_i) | o_i \in children(e) \} $\\
and is $0$ if the node has no children (i.e., it represents a relation).
Note that the 
cost of executing an operation $o$ also takes into account the cost 
of reading the inputs, if they are not pipelined.

Volcano also caches the best plan it finds for each equivalence node, 
in case the node is re-visited during the
depth first search of the DAG. 

A simple extension of the Volcano algorithm to find best plans given
a set of materialized views (in our context, cached results) is 
described in \cite{rssb00:mqo}.  We outline this extension below.

The first step is to identify those equivalence nodes in the query 
DAG  which correspond to cached results.  This is carried out by treating the
cached results as a set of queries, and creating a DAG structure
for the cached results.  The query to be optimized is inserted into
this DAG structure as discussed earlier in Section~\ref{sec:dagrep:queryset}.
Also, as part of this step, subsumption derivations are introduced 
as required.  Thus, given a view $v : \sigma_{A>5}(r)$
and a node (query) $n: \sigma_{A>10}(r)$, a selection operation is inserted
to represent the alternative of computing node $n$ from view $v$.

Let $reusecost(m)$ denote the cost of reusing the materialized 
result of $m$, and let $M$ denote the set of materialized nodes.

To find the cost of a node given a set of nodes $M$ have been materialized, 
we simply use the Volcano cost formulae above for the query, with the 
following change.
When computing the cost of an operation node $o$, if an input equivalence
node $e$ is materialized (i.e., in $M$),  the minimum of 
$reusecost(e)$ and $cost(e)$ is used for $cost(o)$.  
Thus, we use the following expression instead:\\
\hspace*{1cm}
$cost(o) = $ cost of executing $(o)$ + $\Sigma_{e_i \in children(o)}
  C(e_i)$
\vspace{-0.1in}
\begin{tabbing}
xxxxxxxxxxx\=where $C(e_i)$\= \kill
\>where $C(e_i) = cost(e_i)$ if $e_i \not\in M$ \\
\>\>$=min(cost(e_i), reusecost(e_i))$ if $e_i \in M$.\\
\end{tabbing}
\vspace{-0.3in}
Thus, the extended optimizer computes best plans for the query in 
the presence of cached results.  The extra optimization overhead 
is quite small.


\subsections{Representative Set of Queries}
\label{sec:greedy:repset} 

To characterize the dynamic workload,
the cache manager keeps a window of the most recent $N$ distinct
queries as representative of the workload at instant $t$, and
(incrementally) maintains the consolidated query DAG of these queries.
$N$ is a parameter of the algorithm.   This set of $N$ queries is termed the
{\em representative set} and the consolidated query DAG is termed the
{\em representative query DAG} (for the workload) at the instant $t$.

A query may occur more than once in the representative set.
Each query in the representative set 
is weighted by its age -- we  use a weight of $\delta^i$ 
for an occurrence of the query $i$ queries behind the current occurence, 
where $\delta \in (0,1]$ is a decay parameter.  
The overall weight of a query, denoted by
$weight(q)$, is then the sum of the weights of all its occurrences in
the repset.

We shall equate (for description purposes)  a final/intermediate
result with the equivalence node 
in the DAG that represents that result.
Also, for brevity, we shall use the term {\em node} in the rest of the paper
to mean an equivalence node, unless otherwise indicated.

\subsections{Incremental Greedy Algorithm}

We now consider how to decide what results to keep in the cache.
When a query arrives, it is optimized using the current cache
contents, as described in Section~\ref{sec:volc:mat}.

The \incremental\ algorithm then attempts to find out if any of the
nodes of the chosen query plan are worth caching.  These nodes are
available for free, but their benefit must be compared with the benefit
of other nodes.  Specifically, the algorithm described below
compares their benefits only with
nodes that were selected (for caching) when previous queries in the representative set
were considered.

To make the incremental decision, the representative set is updated 
with the new query, and the selection algorithm is applied with 
the candidate nodes being 
\begin{enumerate}
\item nodes selected when the previous query was considered, and
\item all nodes of the chosen query plan.
\end{enumerate}

Suppose a set of nodes $S$ has been selected to be cached.
Given a query $q$, let 
$cost(q,S)$ denote the cost of computing $q$ given that $S$ is in the cache.
Let $cost(R,S)$ be defined as  
\[ cost(R,S) = \Sigma_{q \in R} ~( cost(q,S) \ast weight(q)) \]
i.e., the weighted sum of of the costs of all queries in the 
representative set $R$, given nodes in $S$ are cached.

Given a set of nodes $S$ already chosen to be cached, and a node $x$,
let $benefit(R, x, S)$ be defined as
\[ benefit(R, x, S) = cost(R, S) - (cost(R, \{x\} \cup S) + cost(x, S))
\]
i.e., given nodes in $S$ are already cached, the above formula measures
the additional benefit of caching node $x$.
(In some cases, if $x$ is known to have been computed already,
we will assume $cost(x,S)$ to be $0$; we return to this issue in
later sections.)

\begin{figure}
\begin{small}
\ordinalg{
Procedure {\sc Greedy} \\
{\em Input:} \> \> Expanded DAG for $R$,  the representative set of queries,  \\
\> \> and the set of candidate equivalence nodes for caching \\
{\em Output:} \> \> Set of nodes to be cached \\
\> X = $\phi$ \\
\> Y = set of candidates equivalence nodes for caching  \\
\> while (Y $\neq \phi$) \\
L1:\>\> Among nodes $y \in Y $ such that $size(\{y\} \cup X) < CacheSize)$ \\
  \>\>  \> Pick the node $x $ with the highest $ benefit(R, x, X)/size(x)$ \\
\>\>\>\> /* i.e., highest benefit per unit space */ \\
\> \>  if ($benefit(R,x,X) < 0$) \\
\>\>\>  break; /* No further benefits to be had, stop */ \\
\>\>Y  = Y - x;~~~X = X $\cup$ \{x\} \\
\> return X
}
\end{small}
\vspace{-2mm}
\caption{The Greedy Algorithm}
\label{fig:greedy}
\end{figure}

Figure~\ref{fig:greedy} outlines a greedy algorithm for static caching,
which iteratively picks nodes to be cached.  
Nodes are selected for caching in order of maximum-benefit-first.
It takes as input the set of candidate results (equivalence nodes)
for caching.
At each iteration, among nodes that will fit in the remaining cache space,
the node $x$  that gives the maximum \eat{reduction in the cost} {\em benefit
per unit space}, if it is cached, is chosen to be added to $X$. 
The algorithm  terminates when benefit becomes zero/negative,
or the cache size is exceeded, whichever is earlier.

The output of greedy is a set of nodes that are {\em marked} for caching
and the best plan for the current query.
When the query is executed,  any nodes in its best plan that are marked
are added to the cache, replacing unmarked nodes that are in the cache.

At the end of the incremental greedy selection algorithm, the set of nodes 
selected must fit in cache, but the set may be much smaller than the
cache size. 

Unmarked nodes already in the cache are chosen for replacement on the basis 
of LCS/LRU, i.e. the largest results are preferentially evicted, and 
amongst all results of the same size,  the least recently used one
is evicted.  This policy has been shown to be good by ADMS
\cite{rou:edbt94}.

A variant of the above procedure is to admit even unmarked nodes
in the best plan of the current query in the cache.
We LCS/LRU replacement as before for these unmarked nodes. 
Our performance experiments study the effect of such admission,
and show that it does not provide very large benefits and may be
harmful on the workloads we studied.

\eat{
Thus, LCS/LRU takes care of short to medium-term patterns, 
while our \incremental\ algorithm
caches intermediate results which may be useful for medium
to long-term patterns. 
}

We note that results are cached without any projections, to maximize
the number of queries that can benefit from a cached result.
Extensions to avoid caching very large attributes are possible.

Note that because a query is optimized before fresh caching decisions
are made (see Section \ref{sec:volc:mat})
the chosen best plan for each query is optimal 
for that query, given the current cache contents. 
Plans that increase the cost of a query are not chosen, even if it
generates a node that is beneficial to multiple queries.

%
%

\eat{
Further, unmarked nodes in the best plan of the current query are also
added to the cache on the basis of LCS/LRU.
The decision of what nodes are candidates for the greedy algorithm to
consider for caching had to be made carefully.
The nodes in the best plan of the current query are clearly
candidates for caching.
One option was to treat all nodes that are currently cached 
as candidates.
However, this favored old results over new ones that have a low
benefit, but would be in cache due to recency of usage provided the
cache is not filled with old results chosen by greedy.

We 
Reasons for forming the candidate set
	= nodes in the above plan U nodes marked in the previous invocation
---------------------------------------------------------------------------

Not considering history puts focus on very short-term -- leads to unstable
performance. Too much history goes against any change in the cache.
We consider limited history -- just what is marked in the previous
invocation of greedy.
}

\subsections{Optimizations of Greedy Algorithm}

The greedy algorithm as described above can be expensive due to 
the potentially large number of nodes in the set $Y$,
and the large number of times the function $benefit$ is called,
(which in turn calls the expensive function $cost()$).
Three important optimizations to a greedy algorithm for multi-query
optimization, presented in \cite{rssb00:mqo}, can be adapted for the
purpose of selecting the cachable nodes efficiently:
\begin{enumerate}
\item Consider for caching only
nodes that are shared in some plan for the query, i.e., nodes
that appear as subexpression of two queries, or appear more
than once in some plan for a single query.
(The variable $Y$ in Figure~\ref{fig:greedy} is set to the set of
such nodes.)
We call such nodes {\em sharable nodes}, and for multiquery optimization,
these are the only nodes worth considering for materialization.

Additionally, we may consider nodes for caching even if they are 
not shared, if they are available for free.
For instance, a final/intermediate result of a query can be considered for
caching, since it is available for free, even if it is not shared 
with any other query.
Since some queries may get repeated, it is worth caching final/intermediate 
results of such queries.  That is, if a query is repeated in the repset,
every equivalence node that can form part of a plan for the query is
a candidate (considering only the root of the query is an alternative,
which corresponds, for example, to  \cite{sch96:watchman}).

\item  Since there  are many calls to {\em benefit} (and thereby to
{\em cost()}) at line L1 of Figure~\ref{fig:greedy}, with different parameters,
a simple option is to process each call to $cost$ independent 
of other calls. 
However, observe that the set of cached nodes, which is the second
argument of {\em cost}, changes minimally in successive calls ---
successive calls take parameters of the form $cost(R, \{x\} \cup X)$, 
where only $x$ varies.
That is, instead of considering $x_1 \cup X $ for storing in the cache,
we are now considering storing $x_2 \cup X$ for storing in the cache.
The best plans computed earlier does not change for nodes that are
not ancestors of either $x_1$ or $x_2$.
It makes sense for a call to leverage the work done by a previous call 
by recomputing best plans only for ancestors of $x_1$ and $x_2$.

Our incremental cost update algorithm maintains the state of the DAG (which includes previously 
computed best plans for the equivalence nodes) across calls  
to $cost$, and may even avoid visiting many of the ancestors of 
$x_1$ and $x_2$.

\item With the greedy algorithm as presented above, in each iteration
the benefit of every candidate node that is not yet cached is
recomputed since it may have changed.  If we can assume that the
benefit of a node cannot increase as other nodes are chosen to be
cached (while this is not always true, it is often true in practise)
there is no need to recompute the benefit of a node $x$ if the new
benefit of some node $y$ is higher than the previously computed
benefit of $x$.  It is clearly preferable to cache $y$ at this stage,
rather than $x$ --- under the above assumption, the benefit of $x$
could not have increased since it was last computed.

\end{enumerate}

\subsections{Discussion}
\label{sec:disc}
In addition to the above incremental algorithm,
we considered a cache management strategy that periodically
runs the greedy algorithm with all nodes in the expanded DAG of 
all queries in the representative set.
However, we found the performance of this algorithm to depend  
heavily on the correlation between its period and the periodicity of
queries, leading in many cases to poor gains at a high cost.
We then abandoned this strategy.  In contrast, the \incremental\ algorithm
has a period of 1, but considers a smaller set of nodes for caching.

\sections{Related Work}
\label{sec:relwork}
The closest work related to Exchequer include ADMS  \cite{rou:edbt94},
Dynamat \cite{kot99:dynamat} and  Watchman
\cite{sch96:watchman}.
\eat{
There has been considerable amount of work on query caching in the recent few years.
The work of \cite{dar96:semcache,basu96:predcache} looks at caching 
tuples, but considers only simple selection queries.  
The ADMS system \cite{rou:edbt94} considers select-project-join queries but 
uses simple replacement policies (Least Recently Used, Least Frequently
Used, or Least Cache Size) and does not consider the effect of the 
presence of one result on the benefit of storing another result.

DynaMat \cite{kot99:dynamat} and \cite{des98:chunks} consider only 
caching of cube queries.
In particular, DynaMat focuses on multidimensional range queries which 
can be viewed as a hyper-plane in the Data Cube space. The range 
corresponding to each dimension is either empty, is a single value, 
or is the full range.  Because of this, DynaMat materializes coarser 
grain objects compared to systems that allow arbitrary ranges in 
their view definitions. 

WATCHMAN  \cite{sch96:watchman} considers only caching final query
results, and reuses them only if exactly the same query gets repeated.
It does not consider how to use cached results for other queries.
}
A very recent extension to Watchman \cite{sch99:dyncache} allows 
results to
be used by other queries, but 
(a) restricts the class of queries to select-project-join-aggregate queries, 
and 
(b) caches only selection-free select-project-join-aggregate queries,
i.e., cube queries.


In contrast to ADMS, Dynamat and Watchman, our techniques are
applicable to any class of queries, 
including cube queries.  While cube queries are important, general purpose
decision support systems must support more general queries as well.
Moreover our techniques are extensible in that new operators can 
be added easily, due to the use of the AND-OR DAG framework. 

Query caching systems proposed earlier~\cite{dar96:semcache,
sch96:watchman,rou:edbt94,des98:chunks,kot99:dynamat,sch99:dyncache}, 
maintain statistics for each cached result, which is used to compute 
a replacement metric for the same; 
the replacement metric is variously taken as the cached results
last use, its frequency of use in a given window, its rate of use, etc.
Of the above systems, \cite{sch99:dyncache} and \cite{kot99:dynamat} 
use more sophisticated techniques, specifically computing benefits
of cached results taking other cache contents into account.
However, their techniques are restricted to the case where each
result can be derived directly from exactly one parent (and indirectly
from any ancestor).
Our techniques do not have this restriction.

Moreover, our techniques can find benefits even if the results are not
currently in the cache, and decide to materialize them if they
give an overall benefit, which the other caching techniques are unable
to achieve. 

\eat{
Another closely related area is that of materialized view/index selection.
Materialized view selection can be thought of as a static version of the
caching problem, where decisions are made once, and the cache contents
do not change thereafter (except for keeping them up-to-date in the
presence of updates).  Work in this area includes 
\cite{rss96:matview,venky:sigmod96,venky:index,wisc:matview,sn97:indexsel,lqa97:phys,gupta97:viewsel}.
The algorithms of \cite{venky:sigmod96,gupta97:viewsel} 
use a greedy algorithm to select views with maximum benefit per unit space
(where benefit calculations take into account what has already been
chosen to be materialized).  
Unlike our techniques, 
this body of work typically does not take into account the cost of 
materializing the views selected, and only takes maintenance costs into
account.

A final area of related work is multi-query optimization
\cite{rssb00:mqo,shivku98:transview,arn:ana,cls93:multi,tim:mul,kyu:imp,joo:usi}.
Work in this area attempts to find commonalities between queries
in a given batch of queries, and compute and share common subexpressions to
minimize overall execution cost.  All queries are assumed to be known
a priori, and typically cache space is assumed to be unbounded.

As mentioned earlier, our work exploits the DAG representation and
greedy algorithm presented in \cite{rssb00:mqo}.
However, \cite{rssb00:mqo} does not consider any of the caching issues
discussed here, such as representative sets, periodic recomputation,
when to compute selected views when sufficient slack time is not
available, etc.
}


\sections{Experimental Evaluation of the Algorithms}
\label{sec:perf}

In this section we describe our experimental setup and the results
obtained.

\subsections{System Model}

Our algorithms were implemented on top of the multi-query optimization
code \cite{rssb00:mqo} that we have integrated into our Volcano-based query
optimizer.   The basic optimizer took 
approx.\ 17,000 lines of C++ code, with MQO and caching code taking about 3,000 lines.

\eat{
The optimizer rule set consisted of select push down, join commutativity
and associativity (to generate bushy join trees), and select and
aggregate subsumption.  Our implementation incorporates the
optimizations of \cite{pel97:com} which, for join transformations,
prevent repeated derivations of the same expressions.  Implementation
algorithms included sort-based aggregation, merge join, nested loops
join, indexed join, indexed select and relation scan.
As mentioned earlier, our optimizer handles physical properties 
(sort order and presence of indices) on base and intermediate relations.
}

The block size was taken as 4KB and our cost functions assume 6MB is
available to each operator during execution (we also conducted
experiments with  memory sizes up to 128 MB, with similar
results).  Standard techniques were used for estimating costs, using
statistics about relations.  The cost estimates contain an I/O component
and a CPU component, with seek time as 10 msec, transfer time of 2
msec/block for read and 4 msec/block for write, and CPU cost of 0.2
msec/block of data processed.  We assume that
intermediate results are pipelined to the next input, using an iterator
model as in Volcano;  they are saved to disk only if the result is to be
materialized for sharing.  The materialization cost is the cost of
writing out the result sequentially.

All our cost numbers are estimates from the optimizer.  We validated
our estimates for a few of the queries against actual execution times
on Microsoft SQL-Server 7.0 (SQL queries with hints on execution plan)
and found a good match (within 20 percent).

The tests were performed on a Sun workstation with UltraSparc 10
333Mhz processor, 256MB, and a 9GB EIDE disk, running Solaris 5.7.

\subsections{Test Query Sequences}
\label{sec:workload}

We test our algorithms with streams of 1000 randomly generated queries 
on a TPCD-based star schema similar to the one proposed by 
\cite{sch99:dyncache}. The schema has  a central
{\em Orders} fact table, and four dimension tables 
{\em Part, Supplier, Customer} and {\em Time}. 
The size of each of these tables is
the same as that of the TPCD-1 database. This corresponds to base data
size of approximately 1 GB. Each generated query was of the form:
\begin{quote}
    {\tt
	SELECT SUM(QUANTITY)\\
	FROM ORDERS, SUPPLIER, PART, CUSTOMER, TIME\\
	WHERE {\em join-list} AND {\em select-list}\\
	GROUP BY {\em groupby-list};
    }
\end{quote}

The {\em join-list} enforces equality between attributes of the
order fact table and primary keys of the dimension tables.
The {\em select-list}
i.e., the predicates for the selects were generated by 
selecting 0 to 3 attributes at random from the join result,
and creating equality or inequality predicates on the attributes.
The {\em groupby-list} was generated by picking a subset of \{{\em
suppkey, partkey, custkey, month, year}\} at random.
A query is defined uniquely by the pair {\em (select-list, groupby-list)}.
Even though our algorithms can handle a more general class of queries,
the above class of cube queries was chosen so that we can have a fair
comparison with DynaMat ~\cite{kot99:dynamat} and Watchman
\cite{sch96:watchman}.

There are two independent criteria based on which the pair {\em
(select-list, groupby-list)} was generated:
\begin{itemize}
\item Whether the
queries are:
\begin{itemize}
\item {\em CubePoints:} that is, predicates are restricted to equalities, or
\item {\em CubeSlices:} that is, predicates are a random 
mix of equalities and inequalities.
\end{itemize}
\item The distribution from which the attributes and values are picked
up in order to form the {\em groupby-list} and the predicates in the
{\em select-list}.
\begin{itemize}
\item {\em Uniform Workload:} Uses uniform distribution. All {\em groupby}
combinations and selections are equally likely to occur. 
\item {\em Skewed Workload:} Uses Zipfian distribution with
parameter of 0.5.
The groupby distribution additionally rotates \eat{through the
    groupby-list}
after every interval of 32 queries, i.e.\ the most frequent subset of
groupbys becomes the least frequent, and all the rest shift up one
position.  Thus, within each block of 32 queries, some groupby
combinations and selection constants are more likely to occur than
others.
\end{itemize}
\end{itemize}

Based on the four combinations that result from the above criteria,
the following four workloads are considered in the experiments:
uniform workload of CubeSlices (termed {\em CubeSlices/Uniform}), skewed
workload of CubeSlices (termed {\em CubeSlices/Zipf}),
uniform workload of CubePoints (termed {\em CubePoints/Uniform}),
and skewed workload of CubePoints (termed {\em CubePoints/Zipf}).
Due to lack of space, we present the results only on
CubePoints/Uniform and CubePoints/Zipf.  We did run all our
experiments on CubeSlices/Uniform and CubeSlices/Zipf,
and found their performance is similar to the corresponding CubePoint
queries.

\subsections{Metric} 

The metric used to compare the goodness of caching algorithms is  the
{\em total response time of a set of queries}.  
We report the total response time for a sequence of 900 queries that
enter the system after a sequence of 100 queries  warm up the
cache. This total response time is as estimated by the optimizer and
hence denoted as {\em estimated cost} in the graphs of Section
\ref{sec:exp1}. The estimations are validated by
the results of experiments reported in Section \ref{sec:exp3}.

\eat{
The optimization times were quite low, ranging from 0.03 to 0.3 seconds
per query depending on the caching scheme used.
\reminder{fromKrithi -- need to say what the query execution times were to make
sure people see that the opt times are much lower.}
}

\subsections{List of algorithms compared}

We compare \incremental\ with the following baselines, prior approaches,
and variants.

We consider the following baseline approaches.
\begin{itemize}
\item {\bf NoCache:} Queries are run assuming that there is no
cache. This gives an upper bound on the performance of any well-behaved
caching algorithm.

\item {\bf InfCache:} The purpose of this simulation is to give a lower
bound on the performance of any caching algorithm.  We assume an
infinite cache and {\em do not} include the materialization cost.  Each new
result is computed and cached the first time it occurs, and reused
whenever it occurs later.
\end{itemize}
Experiments are conducted on these two extremes to evaluate the
absolute benefits and competitivity of the algorithms considered.

We evaluate the \incremental\ algorithm against the following prior approaches.
\begin{itemize}
\item {\bf LCS/LRU:} We experiment with the caching policy found to be the
best in ADMS ~\cite{rou:edbt94}, namely replacing the result occupying {\em largest
cache space} (LCS), picking the  {\em least recently used} (LRU) result
in case of a tie. The incoming query is optimized taking the cache contents into
account. The final as well as intermediate results in the best plan are
considered for admission into the cache based on LCS.
It should be pointed out that  LCS/LRU is not aware of the  workload
since each result is assumed independent when caching decisions are made.

\item {\bf DynaMat:}
We simulate DynaMat \cite{kot99:dynamat} by considering only the top-level query results
(in order to be fair to DynaMat,  
our benchmark queries have either no selection or
only single value selections).  The original DynaMat performs matching of
cube slices using R-trees on the dimension space.  In our implementation,
query matching is performed semantically, using our unification
algorithm, rather than syntactically.
We use our algorithms to optimize the query taking into
account the current cache contents; this covers the subsumption
dependency relationships explicitly maintained in\cite{kot99:dynamat}.
The replacement metric is computed as \\
\hspace*{1cm} (number-of-accesses $*$
	cost-of-computation)/(query-result-size)\\
where the number of accesses are from the entire history (observed so far).

\item {\bf WatchMan:}
Watchman \cite{sch96:watchman} also considers caching only the top level query results. The
original Watchman does syntactic matching of queries, with semantic
matching left for future work. We improve on that by considering
semantic matching.  Our Watchman implementation is similar to the
Dynamat implementation described above. The difference is in the
replacement metric: 
instead of using the number of accesses as in the Dynamat implementation, 
we use the rate of use on a window of last ten accesses for each query. 
The replacement metric for Watchman is thus \\
\hspace*{1cm} (rate-of-use * cost-of-computation)/(query-result-size)\\
where the cost of computation is with respect to the 
current cache contents. The original algorithms did not
consider subsumption dependencies between the queries; our implementation
considers aggregation subsumption among the cube queries considered.
\end{itemize}
Given the enhancements mentioned above, the performance of our implementations of
DynaMat and Watchman will be better than the
originally proposed versions. Still, as we will see, the Exchequer
algorithms significantly outperform them.

We consider the following variants of the \incremental\ algorithm.
\begin{itemize}
\item {\bf Incremental/FinalQuery:}
In this variant of \incremental, the caching of intermediate results
is turned off; that is, only the final query result is admitted for each
query in the workload. If the cache is full, a previous unmarked
query result is picked for replacement using LCS/LRU.

\item {\bf Incremental/NoFullCache:}
In this variant of \incremental, only
marked nodes that have been selected from the candidate set
are admitted into the cache. 
If the cache is full, a previous unmarked cached result is
picked for replacement using LCS/LRU.

\item {\bf Incremental/FullCache:}
In this variant of \incremental, apart from the marked nodes selected from
the candidate set, the unmarked
nodes of the query are also considered for admission into the cache.
The marked nodes always get admitted. If the cache is full, a previous
unmarked result is picked for replacement using LCS/LRU. Unmarked nodes
are admitted based on LCS/LRU in the space beyond that used by marked
nodes. The idea is to keep the cache full as
much as possible at all times.
\end{itemize}
The variant that only stores the full query results, but fills up the
cache based on LCS/LRU is not considered. This is because the size of
marked results at any point is expected to be small for our workloads,
and therefore the performance of this variant is expected to be
identical to the LCS/LRU algorithm described above.

As a part of the experimental study below, we evaluate these variants
against each other as well as against the prior approaches.

\eat{
We experiment with different cache sizes:
starting from 10000 to 120,000 blocks.  Note that 10000 blocks
corresponds to 40MB, while 120,000 blocks corresponds to 480MB, which
is about half the size of the 1 GB database.
}

\subsections{Experimental Results}

The size of the representative set is set
to 10 for the experiments. 
We experiment with different cache sizes, corresponding to roughly
0\%, 5\%, 16\%, 32\% and 50\% of the total
database size of approximately 1GB.

\begin{figure}
\centerline{
\psfig{file=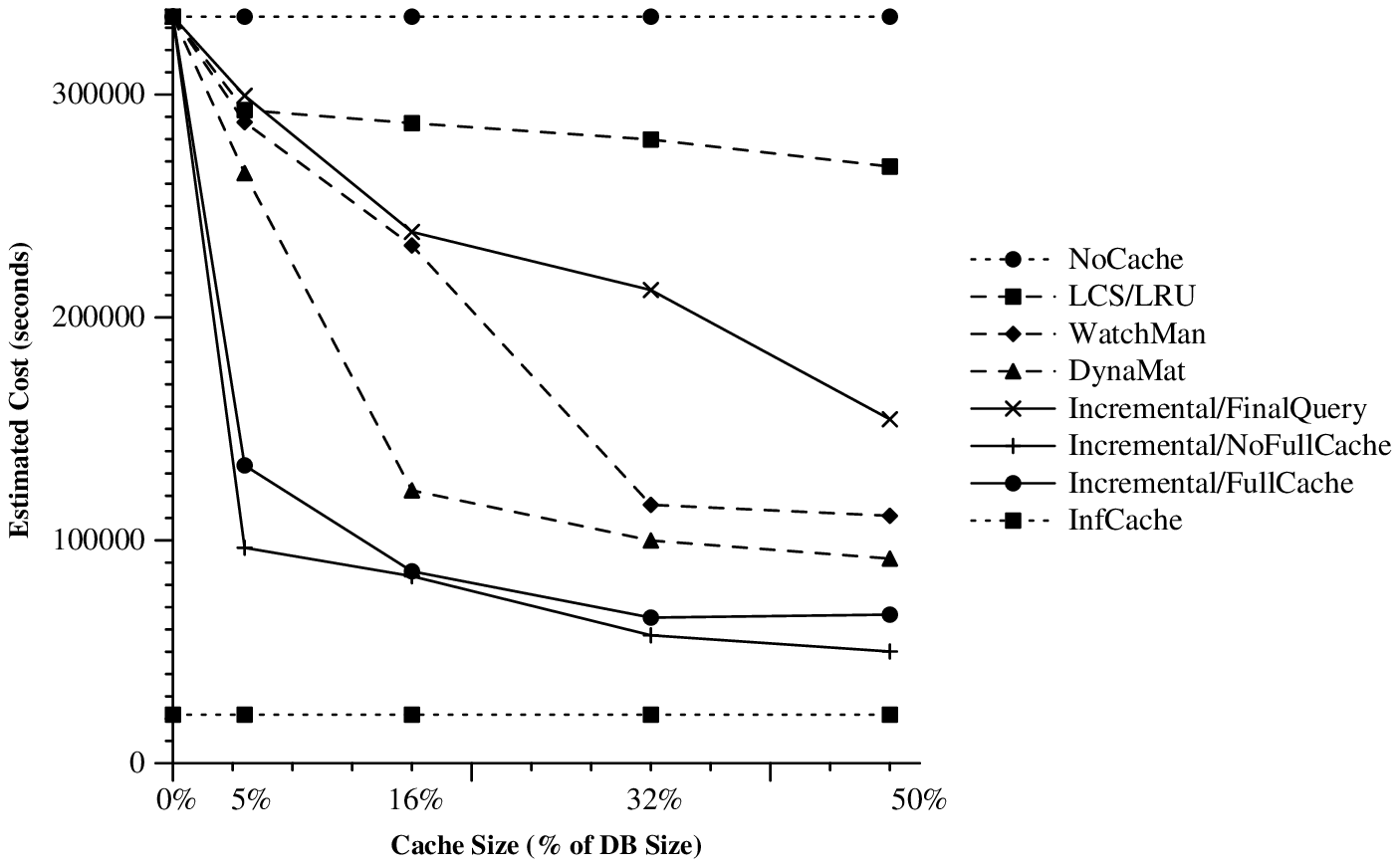,width=5.0in}
}
\caption{Performance on an Uniform 900 Query CubePoints Workload}
\label{fig:cubepoints}
\end{figure}

\begin{figure}
\centerline{
\psfig{file=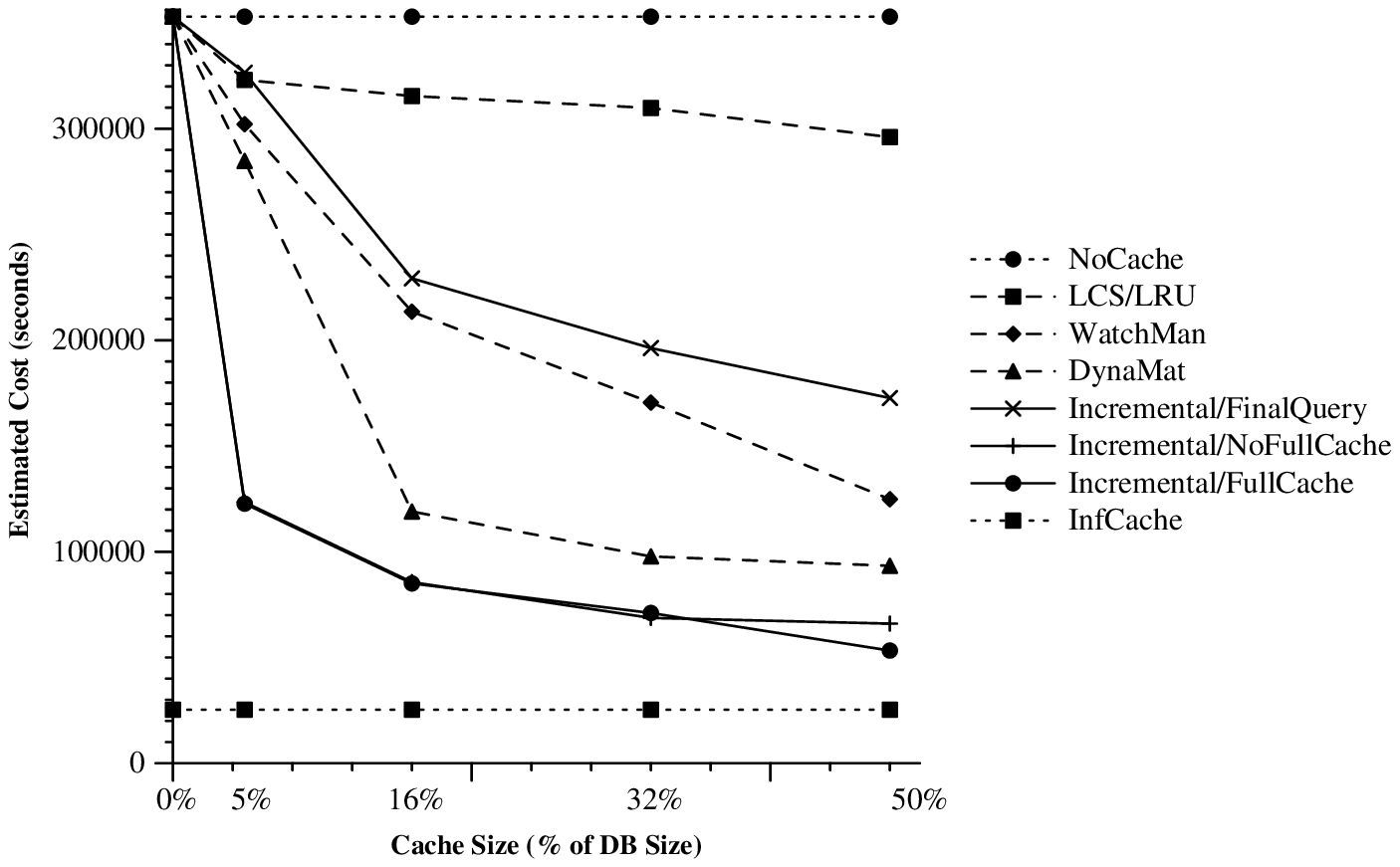,width=5.0in}
}
\caption{Performance on a Zipf/0.5 900 Query CubePoints Workload}
\label{fig:cubepoints-zipf}
\end{figure}

%

\subsubsection {Experiment 1: Comparison of Estimated Response Times}
\label{sec:exp1}
In our first experiment, we compare the performance of
Incremental/FinalQuery, Incremental/NoFullCache and Incremental/FullCache
against each other and 
against \adms, WatchMan and  Dynamat. \nocache\ and \infcache\ are
also included as baselines in order to compare with the worst and best case
performance bounds respectively. The experiments were carried out on the
workloads described in Section~\ref{sec:workload}. 
Figure~\ref{fig:cubepoints} and Figure~\ref{fig:cubepoints-zipf} show the
results of this experiment for CubePoints/Uniform and CubePoints/Zipf
respectively.

\eat{
\adms\ performed worst of the lot, followed by Incremental/FinalQuery
WatchMan, Dynamat and, finally, Incremental/FullCache and
Incremental/NoFullCache. 
}
Incremental/FullCache and Incremental/NoFullCache perform the best
among the algorithms considered, followed by DynaMat, WatchMan,
Incremental/FinalQuery and finally \adms, which performed the worst of
all. 

In all of the considered workloads, Incremental/FullCache and
Incremental/NoFullCache were both much better than the competition for
small cache sizes. As the cache size was increased from 0 to 5\% of the
database size, these algorithms displayed major performance benefits,
almost reaching the performance of WatchMan with much higher cache
sizes.  The better performance of these algorithms can be attributed to
their ability to identify intermediate results that benefit the workload
as a whole.  This is crucial at low cache sizes, where maintaining
intermediate results that benefit most of the queries in the workload
turns out to be a better idea than maintaining a set of overall query
results that can benefit fewer number of queries in the workload due
to recurrence or subsumption.

DynaMat, WatchMan and Incremental/FinalQuery made good improvements as
the cache is enlarged. However, the performance of
Incremental/FinalQuery does not improve further; marked and cached results 
get unmarked after some time and then replacement becomes LCS/LRU, which
is not very effective.
For low cache sizes, the rate of improvement for
DynaMat was much higher than the other two. This can be attributed to
DynaMat having a superior replacement policy which takes the dependency
among the cached results into account. However, the rate of
improvement for DynaMat decreased for larger cache sizes, and the
relative improvement over WatchMan diminished. This is because there is
a lower bound on the response times delivered by algorithms that cache
only the top level results: in case a query cannot be
computed from the result of a previous query, the algorithm pays the
full cost for computing the query (cost of computing from the base
relations).
\eat{
Recall that our implementation of WatchMan exploits the
possibility of exploiting a previously computed result to answer a
query. This is not possible in the original algorithm, and hence the
original algorithm will pay full cost for the first occurrence of each
query, and is thus expected to perform worse than our implementation.
}

\eat{
\adms\ performs very poorly in this experiment. This expected since it
relies on a rule-of-thumb policy, and does not tune its caching policy
intelligently to the workload. As a result, it shows minimal
improvement as cache size is increased.
}

\reminder{check readability}
In Figure~\ref{fig:cubepoints}, observe that as the cache size is
enlarged beyond 32\%, the estimated cost for Incremental/FullCache
increases.  This anomaly can be explained as follows: after a given
cache size (that depends on the workload), the investment made by the
system (in terms of the materialization cost) in caching the extra
results that can be accommodated given the increased cache size does not
pay off in terms of benefits of maintaining these results in the
cache. This behavior for large cache sizes is also displayed
by DynaMat on CubeSlices/Zipf.

Thus, at high cache sizes, the admissibility decisions must be made on
cost-benefit criteria rather than just space availability.
Incremental/NoFullCache scores over Incremental/FullCache in this respect.
Incremental/NoFullCache admits a result into the cache only if it has a
benefit, irrespective of whether there is enough space in the cache; in
contrast, Incremental/FullCache admits a result if there is enough space
in the cache irrespective of its benefit. Thus, as expected,
Incremental/NoFullCache makes more judicious choice for high cache sizes.
In fact, even for small cache sizes, the exclusive cost-benefit criterion can
make an appreciable difference, as is apparent in
Figure~\ref{fig:cubepoints}, where Incremental/NoFullCache
performs better than Incremental/FullCache even for small cache size.

To summarize, {\em these results clearly show the need to cache
intermediate results} in addition to the final results of
queries. Also, even though keeping the cache as full as possible at
all points seems to be a good idea, it is potentially
counterproductive. Overall, they attest to the effectiveness of the
techniques utilized by the Exchequer approach.

\subsubsection{Experiment 2: Space and Time Overheads}
\label{sex:exp2}
The Incremental algorithm is heavily dependent on a multi-query
optimization framework developed in \cite{rssb00:mqo}. This raises
issues about the algorithm's overheads and efficacy in an online
environment. In this section, we address these issues by showing
that the Incremental algorithm has small space and time overheads.

\eat{
Exchequer maintains a consolidated AND/OR DAG of all the queries in the
representative set. As the representative set changes, this DAG is
dynamically maintained by appropriately adding (or deleting)  equivalence
and logical nodes corresponding to the best plans of queries that are
admitted into (or evicted from) the representative set.
}
As an estimate
of the memory overhead the Incremental algorithm, we determined the space
taken by the AND/OR DAG during the execution of the Incremental algorithm.
For the run of Incremental/NoCache on the CubeSlices/Uniform workload,
the DAG took approximately 18M of memory, and was independent of the
cache size.

In terms of the execution time, the optimization of the 1000 queries
comprising the workload took about 20 minutes at cache size of 5\% and
about 48 minutes at cache size of 50\%.  The optimization time
depends on the cache size since the greedy
algorithm chooses nodes only till their size does not exceed the cache
size. Volcano took about 4 minutes to optimize these 1000 queries. This
difference in optimization time is negligible to the total running cost
of the 1000 queries which is of the order of tens of hours as shown by 
our cost estimates, as well as by extrapolation from actual runtimes
of 100 queries on MS SQL Server 7.

\eat{
\begin{verbatim}
sizeof(class Equivalence_t ) = 144
sizeof(class LogicalOp_t ) = 40
sizeof(class PlanGroup_t ) = 172
sizeof(class PhysicalOp_t ) = 28

Numbers for cache size = 5% of db size 
EQNODECOUNT = 8258
LOGNODECOUNT = 41987
PGNODECOUNT = 42011
PHYNODECOUNT = 300977

USRTIME 1174120
SYSTIME 4620

Numbers for cache size = 50% of db size 
EQNODECOUNT = 8177
LOGNODECOUNT = 41862
PGNODECOUNT = 42757
PHYNODECOUNT = 299568

USRTIME 2851440
SYSTIME 4690

no cache
USRTIME 206960
SYSTIME 4430
\end{verbatim}
}

\eat{
\subsubsection{Experiment 3: Validation of the Estimated Response Times}
\label{sec:exp3}
In this experiment, we show that caching intermediate results is
important, and leads to high performance gains.

We construct equi-width histograms with width equal to 1\% of database
size.  We counted the number of distinct intermediate results within
each size group, number of times an intermediate result of a size group
is used, and the total cost savings that result from the same.
Statistics were collected for \incremental\ running on cache with size
50\% of the database size.

The following table gives the results for the {\bf Cubepoints/Uniform}
workload. Cost savings were maximum due to caching intermediate results
in the 3-4\% range.

\begin{tabular}{|l|l|l|l|}
\hline
Result Size &  Result Count & Uses & Cost Savings \\
0-1                 &      2827  & 423  & 14763 \\
1-2                 &      22    & 12   &   777 \\
2-3                 &      23    & 130  & 12557 \\
3-4                 &      3     & 181  & 67255 \\
4-5                 &      5     & 84   & 23148 \\
5-10                &      1     & 44   & 20238 \\ \hline
\end{tabular} \\

The following table gives the corresponding results for the
{\bf Cubepoints/Zipf-0.5} workload.

\begin{tabular}{|l|l|l|l|}
\hline
Result Size &  Result Count & Uses & Cost Savings \\
(\% of db size)  &  & & (seconds) \\ \hline
0-1                  &   2409 & 400  & 15889 \\
1-2                  &   14   & 43   &  5363 \\
2-3                  &   13   & 191  & 12434 \\
3-4                  &   1    & 125  & 77703 \\
4-5                  &   4    & 184  & 54983 \\
5-10                 &   2    & 8    &  2488 \\ \hline
\end{tabular} \\

In both these workloads, hough the number of results as well as the number of uses
were maximum in the 0-1\% range, the maximum cost savings resulted from
the cached results in the 3-4\% range. In fact, both of these have a
bimodal distribution, with peaks occurring at very small result sizes
(less than 1\% of the database size) and again in the 3-4\% size range.

These histograms show that caching performance depends on the
  distribution of sizes of cached results and our algorithms perform
  well since they make intelligent decisions  regarding whether a big
  result should be placed in the cache at the cost
  of a number of smaller results.

Results were similar when the experiment was repeated for smaller
cache sizes.
}

\subsubsection{Experiment 3: Validation of the Estimated Response Times}
\label{sec:exp3}
In this experiment, we study actual run-times of a
sequence of 100 Cubepoint/Uniform queries on Microsoft SQL-Server,
with caching decisions made externally.

The workload was analyzed offline using (a) Incremental/NoFullCache, and
(b) DynaMat (chosen because it is the closest competitor in Experiment
1). The algorithms were instrumented to emit SQL for dynamic
materialization (in the form of temporary tables), indexing and usage of
intermediate results selected for caching by the algorithm. The
temporary tables were created on the {\em tempdb} database in order to
prevent any logging overheads.  The generated SQL script was then
executed as a batch on SQL-Server 7.0 running on Windows NT 4.0 on a 233
MHz Pentium II machine with 128MB of memory. \reminder{check}  The
database size was about 100MB, and the cache size was 4MB.

The following table shows the results of the experiment, as well as the
estimate of the execution cost obtained by \incremental. These results
attest to the efficacy of \incremental, and also verify our cost model.

DynaMat performs badly in this experiment, since there were not many
repetitions of the queries within the stream of 100 queries.
\incremental\ is able to exploit the sharing of intermediate results.

\vspace{2mm}

\begin{tabular}{|l|l|l|}
\hline
Algorithm &  Time on SQL-Server 7.0 & Estimated Response Time \\ \hline
NoCache	              &	46 min &	53 min \\
DynaMat               &	46 min &	51 min \\
Incremental/NoFullCache &	32 min &	31 min \\ \hline
\end{tabular}

\eat{

\sections{Related Work}
\label{sec:relwork}

Our work brings together techniques in query caching,
cache-cognizant optimization, view selection, and multi-query optimization. 
In this section, we review prior work in these areas.

There has been considerable amount of work on query caching in the recent few years.
The work of \cite{dar96:semcache,basu96:predcache} looks at caching 
tuples, but considers only simple selection queries.  
The ADMS system \cite{rou:edbt94} considers select-project-join queries but 
uses simple replacement policies (Least Recently Used, Least Frequently
Used, or Least Cache Size) and does not consider the effect of the 
presence of one result on the benefit of storing another result.

DynaMat \cite{kot99:dynamat} and \cite{des98:chunks} consider only 
caching of cube queries.
In particular, DynaMat focuses on multidimensional range queries which 
can be viewed as a hyperplane in the Data Cube space. The range 
corresponding to each dimension is either empty, is a single value, 
or is the full range.  Because of this, DynaMat materializes coarser 
grain objects compared to systems that allow arbitrary ranges in 
their view definitions. 

WATCHMAN \cite{sch96:watchman} considers only caching final query
results, and reuses them only if exactly the same query gets repeated.
It does not consider how to use cached results for other queries.
A very recent extension of this work \cite{sch99:dyncache} allows 
results to
be used by other queries, but 
(a) restricts the class of queries to select-project-join-aggregate queries, 
and 
(b) caches only selection-free select-project-join-aggregate queries,
i.e., cube queries.


In contrast our techniques are applicable to any class of queries,
including cube queries.  While cube queries are important, general purpose
decision support systems must support more general queries as well.
Moreover our techniques are extensible in that new operators can 
be added easily, due to the use of the AND-OR DAG framework. 

Query caching systems proposed earlier~\cite{dar96:semcache,
sch96:watchman,rou:edbt94,des98:chunks,kot99:dynamat,sch99:dyncache}, 
maintain statistics for each cached result, which is used to compute 
a replacement metric for the same; 
the replacement metric is variously taken as the cached results
last use, its frequency of use in a given window, its rate of use, etc.
Of the above systems, \cite{sch99:dyncache} and \cite{kot99:dynamat} 
use more sophisticated techniques, specifically computing benefits
of cached results taking other cache contents into account.

However, their techniques are restricted to the case where each
result can be derived directly from exactly one parent (and indirectly
from any ancestor).
Our techniques do not have this restriction.

Moreover, our techniques can find benefits even if the results are not
currently in the cache, and decide to materialize them if they
give an overall benefit, which the other caching techniques are not
able to achieve.

Another closely related area is that of materialized view/index selection.
Materialized view selection can be thought of as a static version of the
caching problem, where decisions are made once, and the cache contents
do not change thereafter (except for keeping them up-to-date in the
presence of updates).  Work in this area includes 
\cite{rss96:matview,venky:sigmod96,venky:index,wisc:matview,sn97:indexsel,lqa97:phys,gupta97:viewsel}.
The algorithms of \cite{venky:sigmod96,gupta97:viewsel} 
use a greedy algorithm to select views with maximum benefit per unit space
(where benefit calculations take into account what has already been
chosen to be materialized).  
Unlike our techniques, 
this body of work typically does not take into account the cost of 
materializing the views selected, and only takes maintenance costs into
account.

A final area of related work is multi-query optimization
\cite{rssb00:mqo,shivku98:transview,arn:ana,cls93:multi,tim:mul,kyu:imp,joo:usi}.
Work in this area attempts to find commonalities between queries
in a given batch of queries, and compute and share common subexpressions to
minimize overall execution cost.  All queries are assumed to be known
a priori, and typically cache space is assumed to be unbounded.

As mentioned earlier, our work exploits the DAG representation and
greedy algorithm presented in \cite{rssb00:mqo}.
However, \cite{rssb00:mqo} does not consider any of the caching issues
discussed here, such as representative sets, periodic recomputation,
when to compute selected views when sufficient slack time is not
available, etc.

\sections{Conclusions and Future Work}
\label{sec:concl}

In this paper we have presented new techniques for query result caching,
which can help speed up query processing in data warehouses.
The novel features incorporated in our Exchequer system include optimization aware cache maintenance and the use
of a cache aware optimizer. In contrast, in existing work, the module
that makes cost-benefit decisions is part of the cache manager and
works independent of the optimizer which essentially reconsiders these
decisions while finding the best plan for a query. In our work, the
optimizer takes the decisions for the cache manager. Whereas
existing approaches are either restricted to cube (slice/point)
queries, or  cache just the query results, our work  presents a data-model independent framework and algorithm.
Our experimental results attest to the efficacy of our cache
management techniques and show that over a wide range of parameters
lower query response times (more than 30\% reduction
compared to the best performing competitor) can be achieved.




We have developed several extensions of our techniques, which
we outline below, along with directions for future work.
The first extension is, when we run short of cache space, instead of
discarding a stored result in its entirety, we can discard only parts 
of the result.
For instance, if we have the result of $\sigma_{A<100}(E)$, where
$E$ is some possibly complex expression, we can drop part of the 
stored results and retain $\sigma_{A<10}(E)$ if lower $A$ values
are used more often.  We can implement this by partitioning selection
nodes into smaller selects and replacing the original select by a union
of the other selects.
Two issues in introducing these partitioned nodes are:
(i) What partition should we choose?  and
(ii) If the top level is not a select, we can still choose an attribute 
	to partition on, but which should this be?

The second extension is, instead of discarding a cached result completely,
we can replace it by a summarization.
For example we can replace 
\begin{tabbing}
xxxx\=xxxxx\=\kill
\> {\tt select prod\_type, prod\_id, trans\_id, sales\_amt}\\ 
\> {\tt from sales, product} \\ 
\> {\tt where sales.prod\_id = product.prod\_id}\\
by \\ 
\> {\tt select prod\_type, prod\_id, sales\_amt}\\ 
\> {\tt from sales, product} \\
\> {\tt where sales.prod\_id = product.prod\_id}\\
\> {\tt group by product\_type, prod\_id} 
\end{tabbing}

An important direction of future work is to take updates into
account.  We need to develop techniques for 
(a) taking update frequencies into account when deciding whether 
to cache a particular result, and
(b) decide when and whether to discard or refresh cached results.

We could refresh cached results eagerly as updates happen, or update 
them lazily, when they are accessed.  The results may get discarded
without ever being used, so lazy updates can avoid the cost of
propagating updates in many cases.  
At the time of access we can even choose whether to perform incremental 
update, or recompute the result, or simply discard it and pretend 
it never existed, and use an alternative query plan instead.


\subsections*{Acknowledgements}

We would like to acknowledge the help of Jagadish Rangrej and
Jinesh Vora in implementing part of the code.  

{\small
\bibliographystyle{alpha}
\bibliography{queryopt}
}



\section{Conclusions and Future Work}
\label{sec:concl}

In this paper we have presented new techniques for query result caching,
which can help speed up query processing in data warehouses.
The novel features incorporated in our Exchequer system include optimization aware cache maintenance and the use
of a cache aware optimizer. In contrast, in existing work, the module
that makes cost-benefit decisions is part of the cache manager and
works independent of the optimizer which essentially reconsiders these
decisions while finding the best plan for a query. In our work, the
optimizer takes the decisions for the cache manager. Whereas
existing approaches are either restricted to cube (slice/point)
queries, or  cache just the query results, our work  presents a data-model independent framework and algorithm.
Our experimental results attest to the efficacy of our cache
management techniques and show that over a wide range of parameters
lower query response times (more than 30\% reduction
compared to the best performing competitor) can be achieved. Said
differently, a much smaller cache size (around one tenth) is
sufficient for Exchequer to achieve the same performance as its best
competitor.




We have developed several extensions of our techniques, which
we outline below, along with directions for future work.
The first extension is, when we run short of cache space, instead of
discarding a stored result in its entirety, we can discard only parts 
of the result.
\eat{
For instance, if we have the result of $\sigma_{A<100}(E)$, where
$E$ is some possibly complex expression, we can drop part of the 
stored results and retain $\sigma_{A<10}(E)$ if lower $A$ values
are used more often.
}
We can implement this by partitioning selection
nodes into smaller selects and replacing the original select by a union
of the other selects.
Two issues in introducing these partitioned nodes are:
(i) What partition should we choose?  and
(ii) If the top level is not a select, we can still choose an attribute 
	to partition on, but which should this be?
The second extension is, instead of discarding a cached result completely,
we can replace it by a summarization.
\eat{For example we can replace 
\begin{tabbing}
xxxx\=xxxxx\=\kill
\> {\tt select prod\_type, prod\_id, trans\_id, sales\_amt}\\ 
\> {\tt from sales, product} \\ 
\> {\tt where sales.prod\_id = product.prod\_id}\\
by \\ 
\> {\tt select prod\_type, prod\_id, sales\_amt}\\ 
\> {\tt from sales, product} \\
\> {\tt where sales.prod\_id = product.prod\_id}\\
\> {\tt group by product\_type, prod\_id} 
\end{tabbing}
}
An important direction of future work is to take updates into
account.  We need to develop techniques for 
(a) taking update frequencies into account when deciding whether 
to cache a particular result, and
(b) decide when and whether to discard or refresh cached results.
We could refresh cached results eagerly as updates happen, or update 
them lazily, when they are accessed. 

\eat{
The results may get discarded
without ever being used, so lazy updates can avoid the cost of
propagating updates in many cases.  
At the time of access we can even choose whether to perform incremental 
update, or recompute the result, or simply discard it and pretend 
it never existed, and use an alternative query plan instead.
}


\subsection*{Acknowledgments}

We would like to acknowledge the help of Amlan Haldar, Jagadish Rangrej and
Jinesh Vora in implementing part of the code.  

{\small
\bibliographystyle{alpha}
\bibliography{queryopt}
}
\end{document}